\documentclass[letters, fleqn,usenatbib]{mnras}

\usepackage{newtxtext,newtxmath}

\usepackage[T1]{fontenc}
\usepackage{comment}
\usepackage{bm}
\DeclareRobustCommand{\VAN}[3]{#2}
\let\VANthebibliography\thebibliography
\def\thebibliography{\DeclareRobustCommand{\VAN}[3]{##3}\VANthebibliography}

\usepackage{graphicx,subcaption} 	
\usepackage{amsmath}	

\title[Score-Based Inference of the Early Universe]{Posterior Sampling of the Initial Conditions of the Universe from Non-linear Large Scale Structures using Score-Based Generative Models}

\author[R. Legin et al.]{
Ronan Legin,$^{1,2,3}$\thanks{E-mail: ronan.legin@umontreal.ca}
Matthew Ho,$^{4}$ 
Pablo Lemos,$^{1,2,3,5}$
Laurence Perreault-Levasseur,$^{1,2,3,5,6}$
Shirley Ho,$^{5}$\newauthor
Yashar Hezaveh,$^{1,2,3,5,6}$ and Benjamin Wandelt$^{4,5,7}$
\\
$^{1}$Department of Physics, Universit\'{e} de Montr\'{e}al, Montr\'{e}al, Canada\\
$^{2}$Mila - Quebec Artificial Intelligence Institute, Montr\'{e}al, Canada\\
$^{3}${Ciela - Montreal Institute for Astrophysical Data Analysis and Machine Learning, Montréal, Canada}\\
$^{4}$Sorbonne Universit\'{e}, CNRS, UMR 7095, Institut d'Astrophysique de Paris, 98 bis bd Arago, 75014 Paris, France\\
$^{5}$Center for Computational Astrophysics, Flatiron Institute, 162 5th Avenue, 10010, New York, NY, USA\\
$^{6}$Perimeter Institute for Theoretical Physics, Waterloo, Ontario, Canada, N2L 2Y5\\
$^{7}$Sorbonne Universit\'{e}, Institut Lagrange de Paris, 98 bis boulevard Arago, 75014 Paris, France}

\date{Accepted XXX. Received YYY; in original form ZZZ}

\pubyear{2023}

\begin{document}
\label{firstpage}
\pagerange{\pageref{firstpage}--\pageref{lastpage}}
\maketitle

\begin{abstract}
Reconstructing the initial conditions of the universe is a key problem in cosmology. Methods based on simulating the forward evolution of the universe have provided a way to infer initial conditions consistent with present-day observations. However, due to the high complexity of the inference problem, these methods either fail to sample a distribution of possible initial density fields or require significant approximations in the simulation model to be tractable, potentially leading to biased results. In this work, we propose the use of score-based generative models to sample realizations of the early universe given present-day observations. We infer the initial density field of full high-resolution dark matter N-body simulations from the present-day density field and verify the quality of produced samples compared to the ground truth based on summary statistics. The proposed method is capable of providing plausible realizations of the early universe density field from the initial conditions posterior distribution marginalized over cosmological parameters and can sample orders of magnitude faster than current state-of-the-art methods.
\end{abstract}



\section{Introduction}

In the standard model of cosmology, structure originates from quantum fluctuations of a primordial density field, which are scaled to macroscopic distances by a physical process called inflation~\citep{guth1981inflationary, albrecht1982cosmology, linde1982new, linde1983chaotic}. This initial density field represents the seed to all structure seen in the Universe today. 
Furthermore, different models of inflation predict various levels of non-Gaussianity in the primordial density field~\citep{acquaviva2002second, maldacena2003non, bartolo2004non}. Therefore, accurate methods to reconstruct this primordial field could shed light on the unknown mechanism behind inflation and guide our search for new physics. 

Beyond the study of early universe physics, knowledge of the initial conditions of the Universe can be combined with a forward model to  compute predictions for any observable on our past lightcone. Such predictions can act as discovery templates for new physical effects in cross-correlation with external data, e.g., for the discovery of secondary anisotropies in the microwave sky;  be used to perform posterior predictive tests of the underlying cosmological physics model; or give observational access to quantities that were hitherto only accessible in simulations, such as the dynamical assembly history of elements of the cosmic web, such as clusters, filaments, or voids \citep{2010MNRAS.403.1392L,2015JCAP...06..015L,jasche2019physical,https://doi.org/10.48550/arxiv.2212.07840}. For these reasons, substantial effort has been put in the inference of initial conditions as one of the key problems in cosmology. Currently, the primary constraints on the initial conditions come from linear reconstruction \citep{Komatsu2005, Yadav2005} applied to observations of the Cosmic Microwave Background \citep[CMB;][]{2020A&A...641A...1P, akrami2020planck}.

Upcoming galaxy surveys will provide vast amounts of information on small scales, begging the question: can we infer the small-scale initial conditions from non-linearly evolved structure? This inverse problem compounds the challenges of high dimensionality (representing the initial density field on a 3-D grid of $10^7$ voxels corresponds to a $10^7$ dimensional inference problem) and the complexity of computing the non-linear forward mapping between the primordial initial conditions and the present-day density field, requiring, at the very least, high-resolution N-body simulations~\citep{springel2005cosmological, villaescusa2020quijote} and further modeling of the distribution of observable tracers of the underlying dark matter field \citep{2015ARA&A..53...51S}.

Current state-of-the-art methods such as Bayesian Origin Reconstruction of Galaxies \citep[BORG;][]{Jasche2013, jasche2019physical} use Hamiltonian Monte Carlo, a Markov Chain Monte Carlo (MCMC) algorithm for Bayesian parameter inference of the initial conditions, in which one can exploit the gradient of the likelihood to efficiently generate posterior samples. However, due to the high computational cost of full N-body simulations, BORG relies on approximate simulation methods such as second-order Lagrange Perturbation Theory (2LPT) and Particle-Mesh (PM) simulations, which are inaccurate at small scales. Moreover, they require fully-differentiable simulators, which has so far precluded including non-differentiable operations---such as halo finding---that are part of standard modeling pipelines that map dark-matter density fields to galaxy surveys.

In the hopes of improving upon these limitations, machine learning has been used to reconstruct the initial conditions from simulations \citep{Modi2021, 2023MNRAS.520.6256S}. Unfortunately, due to the high-dimensional parameter space, these works have been limited to predicting a single point-estimate, as modeling the full multi-million-dimensional posterior distribution is intractable. Since these models do not produce samples of the early universe, they do not provide any measures of uncertainty on the reconstructions. 

In this work, we propose the use of score-based generative diffusion models \citep{Song2019, Ho2020, Song2020} to learn the distribution of early universe density fields conditioned on the present-day matter density field and to produce samples from it. We train a neural network to predict the score of the posterior distribution using simulations from the Quijote Latin Hypercube set \citep{2020ApJS..250....2V}. We then use the estimate of the score network to solve a reverse-diffusion stochastic differential equation to sample the posterior distribution of the initial conditions.

\section{Methods}
\subsection{Problem Overview}

Our goal is to infer the 3-D density field of the early universe $\bm{x}$ given observations $\bm{y}$ of the dark matter distribution at low redshift. We can define this problem within a Bayesian framework, where we are interested in sampling from the posterior distribution $p(\bm{x}|\bm{y})$. Using Bayes's theorem, the posterior distribution $p(\bm{x}|\bm{y})$ can be written as,
\begin{equation}
    p(\bm{x}|\bm{y}) = \frac{p(\bm{y}|\bm{x})p(\bm{x})}{p(\bm{y})},
\end{equation}
where $p(\bm{y}|\bm{x})$, $p(\bm{x})$, $p(\bm{y})$ represent the likelihood, prior and evidence respectively. The prior distribution $p(\bm{x})$ reflects our knowledge on the possible realizations of the early universe, while the evidence $p(\bm{y})$ gives the probability for a realization of the data. The likelihood distribution $p(\bm{y}|\bm{x})$ represents the distribution of possible observations $\bm{y}$ given a fixed realization of the early universe $\bm{x}$. It includes our cosmological forward simulator and additional effects that resemble true observations (e.g., selection functions, galaxy shot noise). For our problem, we let $\bm{y}$ be the simulated present-day comoving dark matter density field with added noise sampled from a normal distribution with standard deviation of 0.1. 
As detailed in section \ref{sims}, the early and late density fields, $\bm{x}$ and $\bm{y}$, are represented on 3-D mesh grids with $128^3$ voxels. Therefore, the sampling space of $\bm{x}$ is multi-million in dimension. Because of the high dimensionality of the inference problem, modeling the posterior using density estimation techniques such as normalizing flows is not feasible. 

Instead of directly modelling the posterior $p(\bm{x}|\bm{y})$, we can model the gradient of the log posterior distribution $\nabla_{\bm{x}} \log p(\bm{x}|\bm{y})$. In comparison, $\nabla_{\bm{x}} \log p(\bm{x}|\bm{y})$ is computationally tractable, as it does not depend on the normalization of the posterior distribution. This means that it can be approximated by a simple neural network that learns a function $s(\bm{x}, \bm{y})$ mapping a set of inputs ($\bm{x}$, $\bm{y}$) to an output prediction of the score $\nabla_{\bm{x}} \log p(\bm{x}|\bm{y})$. In the following section, we describe how we can train a conditional neural network to estimate the score $\nabla_{\bm{x}} \log p(\bm{x}|\bm{y})$ and use it in the context of score-based generative models \citep{Song2019, Song2020} to produce samples from the posterior distribution $p(\bm{x}|\bm{y})$.

\begin{figure}
    \centering
    \includegraphics[width=0.5\columnwidth]{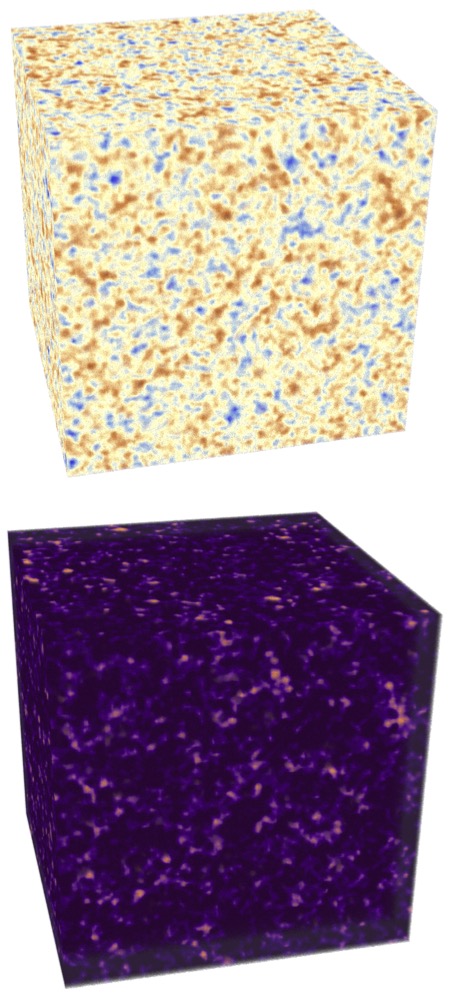}
    \caption{Top: Early universe dark matter density field at redshift $z = 127$. Bottom: Present-day dark matter density field at redshift $z = 0$. The goal is to infer the early universe (top cube) given the present-day dark matter density field (bottom cube). Each face of the cubes shows the outermost pixels for a $(1\ \text{Gpc}/h)^3$ simulation volume. The density cubes are discretized on a $128^3$ grid.}
    \label{cubes}
\end{figure}

\subsection{Score-Based Generative Models}
Score-based generative modeling is a framework designed to learn the distribution of variables from a dataset, by approximating the score $\nabla_{\bm{x}} \log p(\bm{x})$, which is typically modeled by a neural network \citep{Song2020}. The procedure also provides the possibility to generate new samples from the learned distributions. They have been used to great success across a wide variety of domains \citep[e.g.,][]{2021arXiv211108005S, 2021arXiv210506337P, Gnaneshwar2022, Anand2022, 2022arXiv221103812A, 2022arXiv221112444M, 2023arXiv230203046L}, and have surpassed previous state-of-the-art methods such as GANs \citep[e.g.,][]{2021arXiv210505233D, 2022arXiv221207501M}. These achievements have been made possible due to a number of technical improvements in regard to the sampling strategy used to generate samples. In part, one important breakthrough consisted of framing the sampling method as a reverse-diffusion process, where the data $\bm{x}$ is perturbed at various noise levels, and sampling new data points $\bm{x}$ consists of iteratively reversing this process starting from pure noise. In \cite{Song2020}, this process is defined as the stochastic differential equation (SDE):
\begin{equation}
\label{fwdsde}
    d\bm{x} = f(\bm{x}, t) dt + g(t) d\bm{w},
\end{equation}
where $f(\bm{x}, t)$ is called the \textit{drift} term, $d\bm{w}$ is a Wiener process characterizing the random noise, and $g(t)$ is a scalar function that determines the level of added noise. The key to generating samples $\bm{x}$ from $p(\bm{x})$ lies in reversing the diffusion process by solving the reverse-SDE:
\begin{equation}
\label{rvdsde}
    d\bm{x} = \left(f(\bm{x}, t) - g(t)^2 \nabla_{\bm{x}} \log p_{t}(\bm{x}) \right)dt + g(t) d\bm{w},
\end{equation}
where $p_{t}(\bm{x})$ is the probability distribution of $\bm{x}$ at time $t$. Alternatively, we can extend the previous equation to reverse a conditional diffusion process by solving backward in time
\begin{equation}
\label{rvdsde_cond}
    d\bm{x} = \left(f(\bm{x}, t) - g(t)^2 \nabla_{\bm{x}} \log p_{t}(\bm{x}|\bm{y}) \right)dt + g(t) d\bm{w},
\end{equation}
which requires the conditional score $\nabla_{\bm{x}} \log p_{t}(\bm{x}|\bm{y})$ at time $t$.

In this work, we train a neural network conditioned on the observation $\bm{y}$ and time $t$, denoted $s(\bm{x},\bm{y}, t)$ to learn the score of the posterior, $\nabla_{\bm{x}} \log p_{t}(\bm{x}|\bm{y})$, via score matching \citep{JMLR:v6:hyvarinen05a, 10.1162/NECO_a_00142, Song2020}. We then solve Equation \ref{rvdsde_cond} to sample from the posterior distribution of initial conditions $p(\bm{x}|\bm{y})$ by replacing the score $\nabla_{\bm{x}} \log p_{t}(\bm{x}|\bm{y})$ by its approximation from the score network, $s(\bm{x},\bm{y},t)$.

\subsection{Solving the Reverse-SDE}

There exist many numerical methods to solve the reverse-SDE from Equation \ref{rvdsde_cond}. We choose the Euler-Maruyama method which discretizes the SDE as

\begin{equation}
    \label{euler_maruyama}
    \bm{x}_{t + \Delta t} = \bm{x}_{t} + \left( f(\bm{x}_{t},t) - g(t)^2 \nabla_{\bm{x}_t} \log p_{t}(\bm{x}_t|y) \right) \Delta t + g(t) \bm{z}_t \sqrt{-\Delta t},
\end{equation}

where $\Delta t = -1/N$ is the step size, $N$ is the number of steps and $\bm{z}_{t}$ is sampled from a standard normal distribution with the same dimensions as $\bm{x}_{t}$. We perturb the initial conditions $\bm{x}$ with noise following the Variance Exploding Stochastic Differential Equation (VESDE) proposed in \cite{Song2020}. We then solve the discretized reverse-SDE from Equation \ref{euler_maruyama} to sample from the posterior distribution $p(\bm{x}|\bm{y})$. Based on the geometric interpretation from \citet{song2020improved}, we choose a maximum and minimum diffusion noise level of $\sigma_{\text{max}} = 100$ and $\sigma_{\text{min}} = 0.01$ respectively, and choose $N = 1000$ steps.

\subsection{Score Network Architecture and Training}
The score network architecture is based on the \textit{Pytorch} implementation by \cite{Song2020} available on GitHub\footnote{\url{https://github.com/yang-song/score_sde_pytorch}}.
The network follows the RefineNet architecture \citep{2016arXiv161106612L} with five downsampling and upsampling levels each with 2 \textit{ResNet} blocks from the \textit{BigGAN} model \citep{Brock2018} and with outputs of 32 features maps for the first, fourth and fifth level and 64 feature maps for the second and third level. Each \textit{ResNet} block contains a \textit{Dropout} layer \citep{Dropout2014} with a dropout rate of $10\%$ during training. The main difference with the implementation by \cite{Song2020} is that $\bm{y}$ is concatenated to $\bm{x}_t$ along the channel dimension and fed as input to the network. A proof showing the validity of this input scheme to learn the score of conditional probability distributions can be found in \cite{2021arXiv211113606B}.
We train the network for approximately 400 epochs with a batch size of 8 split across 4 NVIDIA H100 80GB Graphical Processing Units (GPU). We use the Adam optimizer with a learning rate of $2 \times 10^{-4}$ and clip the gradient of the weights to a maximum gradient norm of 1. The duration of training under these settings is approximately 24 hours.

\subsection{Simulations}
\label{sims}
We use density fields from the Quijote Latin Hypercube set of N-body simulations \citep{2020ApJS..250....2V} to train and test the score network. Specifically, we use the set of 2,000 $(1\ \text{Gpc}/h)^3$ dark matter N-body simulations without massive neutrinos. The simulations initialized with different random seeds run with different values for the cosmological parameters in the range $\Omega_m \in [0.1, 0.5]$, $\Omega_b \in [0.03, 0.07]$, $h \in [0.5, 0.9]$, $n_s \in [0.8, 1.2]$ and $\sigma_8  \in [0.6, 1.0]$. We use 1,900 N-body simulations for training and 100 for testing with the $\bm{x}$ and $\bm{y}$ density fields computed on a $128^3$ grid at redshift $z = 127$ and redshift $z = 0$ respectively. An example of initial conditions and present-day density field is shown in Figure \ref{cubes}. In this work, the redshift $z = 127$ density fields we train the score network with are individually normalized by their own variance, resulting in generated posterior samples of the early universe density field with variance of one. We found that this improves training convergence and in practice, the generated samples can be rescaled with the true variance predicted from the redshift $z = 0$ density field power spectrum from e.g. CAMB \citep{2000ApJ...538..473L, 2020ApJS..250....2V}. Note that the samples predicted at redshift $z = 127$ are within the linear regime on the scales we consider and therefore can be related directly to an earlier density field (e.g., $z \sim 1000$) using linear theory.

\section{Results}

\begin{figure*}
    \centering
    \includegraphics[width=\textwidth]{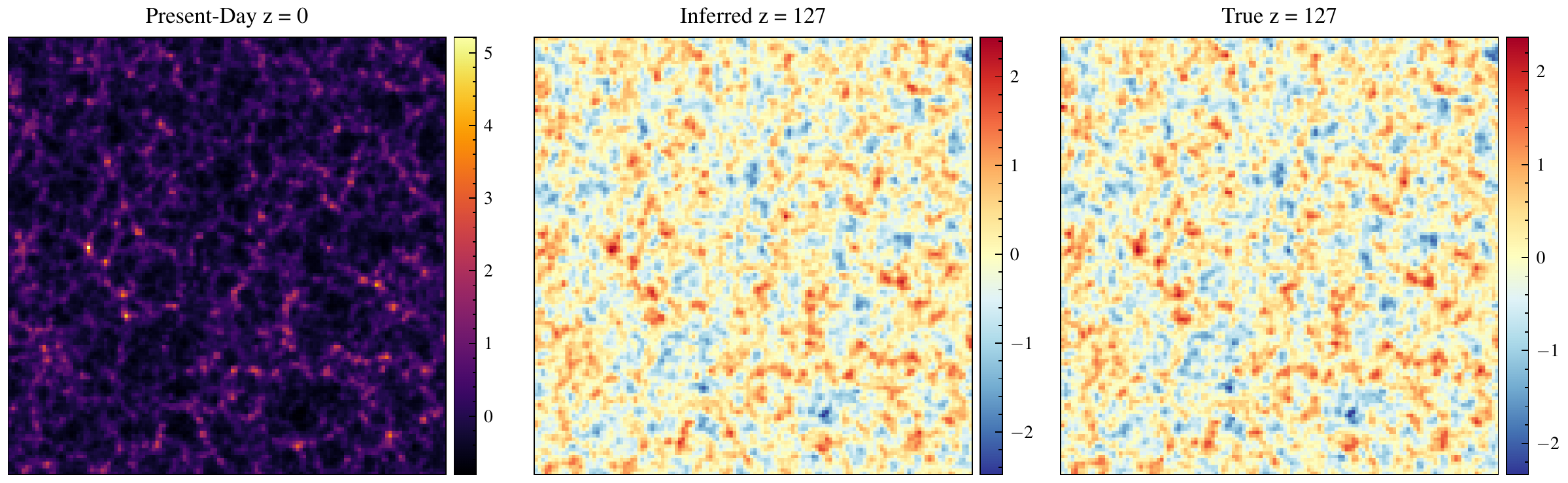}
    \caption{Left: The density field at redshift $z = 0$ for the fiducial Planck cosmology. Center: Initial conditions sampled from the posterior $p(\bm{x}|\bm{y})$. Right: The true initial conditions. All three density fields span a $1000 \times 1000 \times 125 (h^{-1} \text{Mpc})^3$ region averaged over the third axis. This example demonstrates the capability of score-based generative models to sample highly detailed initial conditions consistent with the ground truth. See Figure \ref{std_fid} for quantification of uncertainty.}
    \label{obs_sample_true}
\end{figure*}

\begin{figure}
    \centering
    \includegraphics[width=0.8\columnwidth]{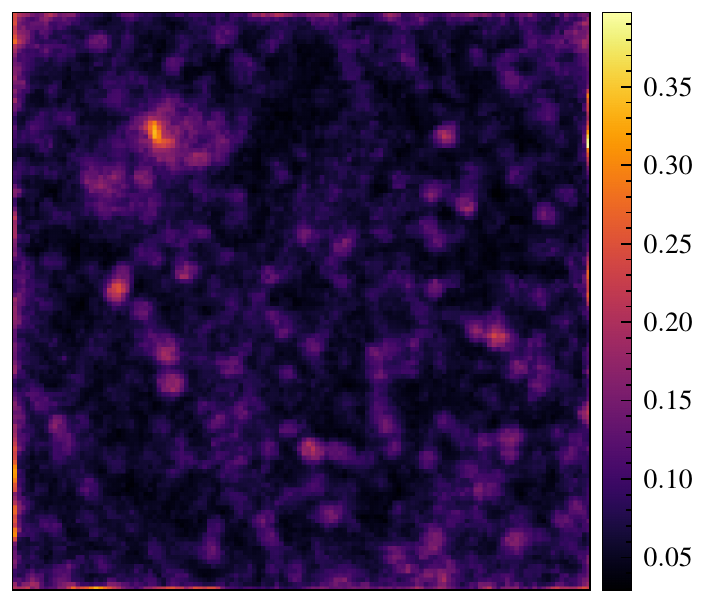}
    \caption{Posterior variance of initial condition samples from the fiducial Planck cosmology simulation per pixel averaged over the depth of a volume of $1000 \times 1000 \times 125 (h^{-1} \text{Mpc})^3$. Reconstruction variance is high in patches that will collapse into large halos at the present day (see Figure \ref{obs_sample_true}). 
    As expected, posterior variance also increases near the boundaries of the data volume.}
    \label{std_fid}
\end{figure}

We show results using our score network on density fields from the fiducial Quijote set of simulations using the fiducial Planck cosmology with $\Omega_m = 0.3175$, $\Omega_b = 0.049$, $h = 0.6711$, $n_s = 0.9624$ and $\sigma_8  = 0.834$, and from six of the Quijote Latin Hypercube set of simulations from our test set with different cosmological parameter values. Note that our score network was not trained on density fields from fiducial cosmology simulations. For each simulation, we solve Equation \ref{rvdsde_cond} to produce 100 samples from the posterior distribution $p(\bm{x}|\bm{y})$ using our trained score network $s(\bm{x}, \bm{y}, t)$. In Figure \ref{obs_sample_true} and in appendix \ref{appendixA}, we show examples of sampled initial conditions for a simulation with the fiducial cosmology. In Figure \ref{std_fid}, we show a map of the posterior sample variance based on the same present-day density field shown in Figure \ref{obs_sample_true}. We then compare the power spectrum of the predicted samples and ground truth, compute the transfer function and cross-correlation, and show the results in Figure \ref{pspec_fid} for the fiducial cosmology case and in appendix \ref{appendixA} for the latin-hypercube simulations.

\begin{figure}
    \centering
    \includegraphics[width=0.9\columnwidth]{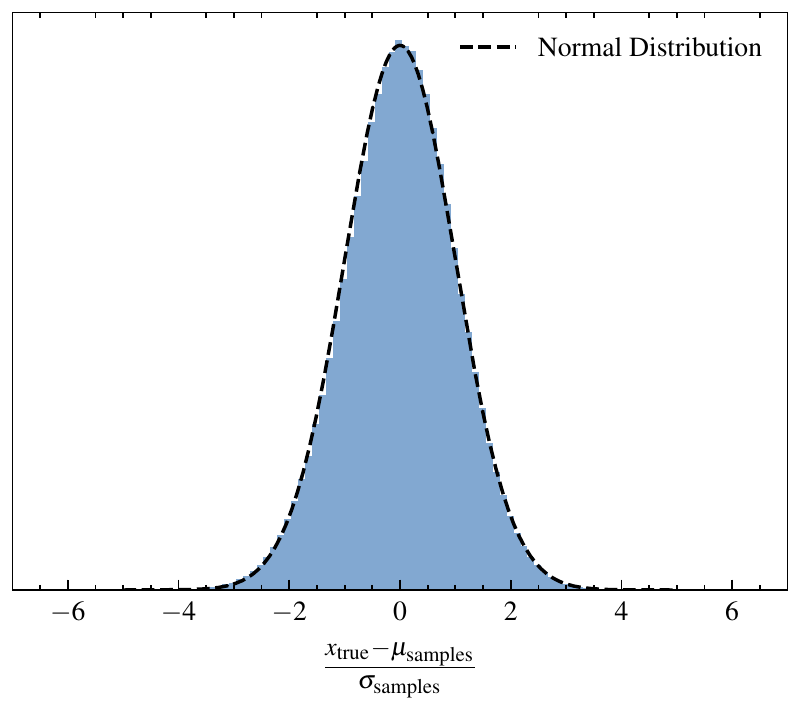}
    \caption{Posterior coverage test for the fiducial Planck cosmology: we show the histogram of the difference between the true initial condition $\bm{x}_{\text{true}}$ and the posterior mean $\mu_{\text{samples}}$ normalized by the posterior standard deviation $\sigma_{\text{samples}}$. If the inferred posterior samples of initial conditions have the correct mean and variance of the voxel-wise marginal posterior distributions, we expect this histogram to have mean 0 and standard deviation 1.
    \label{cov_fid}
    }
\end{figure}

Moreover, we perform a posterior coverage test to verify if the distribution of samples accurately cover the true posterior distribution. This consists of computing the difference between the true initial conditions and the mean of the initial conditions posterior samples, and then dividing by the standard deviation of the posterior samples. If the distribution of the resulting values is consistent with the statistics of a normal distribution (mean of zero and standard deviation of one), then the generated posterior samples have the correct first two marginal moments of the true posterior distribution $p(\bm{x}|\bm{y})$. We show results of this coverage test in Figure \ref{cov_fid} for the fiducial simulation.

\begin{figure}
    \centering
    \includegraphics[width=\columnwidth]{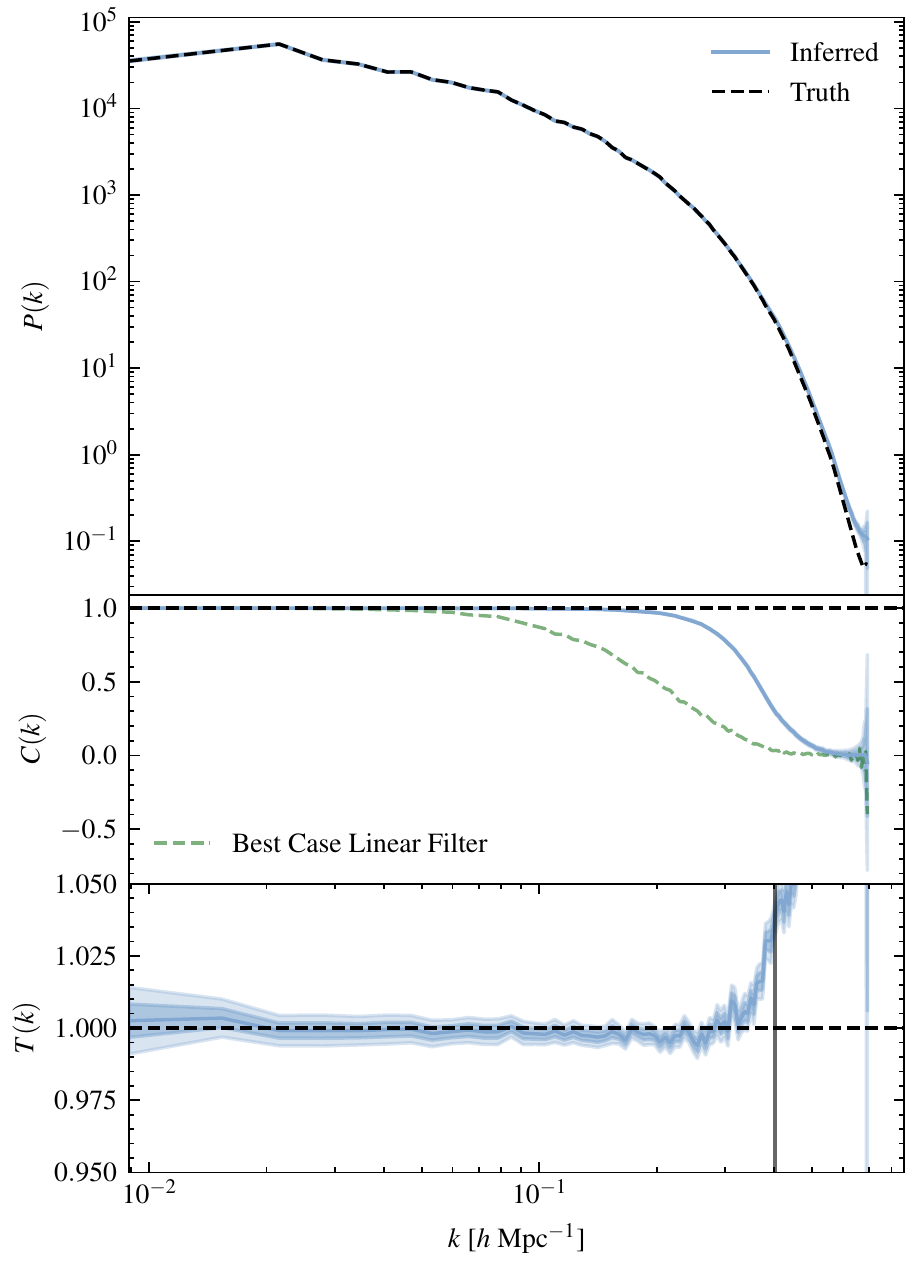}
    \caption{Test statistics to verify the accuracy of samples from the posterior distribution of initial conditions against the ground truth from the fiducial Planck cosmology simulation. Note that this cosmology was not used for training the score network. Top: The power spectrum of the redshift $z = 127$ density field vs the true density field. Middle: The cross-correlation $C(k)$ between every posterior sample and the true field. For comparison we show same for the data and the true field, corresponding to the best achievable $C(k)$ for an optimal linear filter (green dashed). The posterior samples contain significantly more information. Bottom: The transfer function between the posterior samples and the true field with the Nyquist frequency shown as the vertical black line. The shaded regions represent $1\sigma$ and $2\sigma$ errors. The results illustrate the high level of accuracy of the inferred initial conditions.}
    \label{pspec_fid}
\end{figure}

\section{Discussion}
In this work, sampling possible initial conditions from the multi-million-dimensional posterior distribution is done by solving a diffusion process backwards in time. This requires the score $\nabla_{\bm{x}} \log p_t(\bm{x}|\bm{y})$, which guides the diffusion towards the posterior distribution while random noise diffuses the samples in order to explore the sampling space. Compared to the posterior probability $p(\bm{x}|\bm{y})$, modelling $\nabla_{\bm{x}} \log p_t(\bm{x}|\bm{y})$ is computationally tractable as it is independent of the normalization of $p(\bm{x}|\bm{y})$, which is intractable given the high dimensionality of the inference problem. Therefore, methods that directly learn the posterior density using models such as masked autoregressive flows \citep{2017arXiv170507057P} are not suitable for this task.

Efficient MCMC sampling methods in high dimensions such as Hamiltonian Monte Carlo (HMC), as implemented in BORG  \citep{Jasche2013,jasche2019physical}, also do not need a normalized posterior distribution, but they do require a differentiable forward model that must be run for each step in the integration of the Hamiltonian trajectory. For our problem, each forward step amounts to running an N-body simulation from initial conditions until the present day. Each such run must be combined with an adjoint run to compute the gradient. Therefore, the generation of each new sample from the posterior requires tens of simulations. As in all MCMC techniques, accepted samples are correlated leading to a burn-in stage that limits the parallelism of the approach since each running chain must first converge to the posterior distribution before generating useful samples. Moreover, the effective number of samples in the chain is smaller than the number of accepted samples by a factor proportional to the correlation length of the chain. As a result BORG must run on the order of 100 N-body simulations for each approximately independent sample drawn from the posterior distribution, although this could be improved by fitting a variational approximation for the MCMC proposal distribution \citep{modi2023reconstructing}. Because of this, BORG uses approximate N-body simulators such as PM, which sacrifice accuracy for speed, as they are typically orders of magnitude faster than full N-body simulations. Even so, it can require upwards of 1000 CPU hours to generate a single independent posterior sample of initial conditions. Given this, using a full N-body simulator as the forward model for BORG is not computationally feasible. 

In contrast, the method used in this work performs inference using full N-body simulations and can generate independent samples of initial conditions from the posterior distribution within minutes on a single GPU. The sampling procedure can also be trivially parallelized by independently solving Equation \ref{rvdsde_cond} across multiple devices. Furthermore, our approach samples from the posterior distribution of initial conditions marginalized over cosmological parameters. There are straightforward generalizations to score-based sampling that will allow sampling from the joint posterior distribution of initial conditions and cosmological parameters. This will be explored in future work once we move to more realistic models of the observables.

We now make a qualitative assessment of whether our posterior samples exploit all the available information. As our work is the first to sample from the posterior distribution of initial conditions using full N-body simulations, it opens the door to using the full information content of future sky surveys for inference.  Figure \ref{pspec_fid} demonstrates that the posterior samples reproduce the true power spectrum to an accuracy of better than 1\% for all $k$ except within <10\% of the Nyquist frequency of the grid. The reconstructed fields account for 92\% of the variance of the true initial field; this level is expected given that gravitational collapse destroys information within Lagrangian patches that fall into halos. Using the Press-Schechter formalism and cutting off the power spectrum at the grid  frequency  gives an estimated collapse fraction of 12\% in resolved halos. This is close to the measured fraction in the simulation of 16\%. These calculations suggest that our samples  saturate the information that is physically available given non-linear gravitational collapse and coarse-graining on the grid scale.

It bears mentioning that our analysis of DM density fields represents a more stringent test of the network capability than what will be required for the analysis of realistic observations, since these are far more sparsely sampled and hence noisier than the DM density field. 

A possible application of our methodology would be to run N-body simulations from the samples of the inferred initial conditions; for existing samplers, such as BORG, this is a natural byproduct \citep{2017JCAP...06..049L}. These constrained realisations would open the possibility towards unlocking important cosmological questions; we would obtain samples from the posterior distribution of possible N-body simulations that result in the final conditions. This would allow us to sample the posterior over properties of galaxy clusters, such as the positions, masses, and velocities of all dark matter halos in surveys. Furthermore, the possibility to sample constrained high resolution N-body simulations would provide us with the means to uncover the set of possible halo assembly histories. For example, this could be used to sample possible histories of the Local Group and the Milky Way, as in \cite{sibeliusproject}. Initial condition inference with full N-body simulations would help ensure accuracy since galaxy assembly histories are sensitive to scales that are deeply in the non-linear regime today.

\section{Conclusions}
This work proposes score-based generative models for efficient sampling of the posterior distribution of initial conditions, a problem that has up until now been intractable using high-resolution N-body simulations. The key is to sample by solving a reverse-diffusion process requiring only the gradient of the noise-perturbed log posterior $\nabla_{\bm{x}} \log p_{t}(\bm{x}|\bm{y})$, which  can be learned using neural networks. The results show that we can perform accurate inference of the initial conditions marginalized over cosmological parameters at a fraction of the cost of state-of-the-art methods, generating samples from the posterior within minutes on a single GPU. Our tests indicate that the samples have the correct statistics at a level of better than 1\% across the relevant range of scales. In future work, we aim to expand the proposed approach to infer initial conditions from simulated halo and galaxy catalogs and ultimately apply it to real data. 

\section*{Acknowledgements}

This work is supported by the Simons Collaboration on ``Learning the Universe". The Flatiron Institute is supported by the Simons Foundation. This work is also supported by the NASA Research Opportunities in Space and Earth Science (ROSES) program through grant number 12-EUCLID12-0004. The work is in part supported by computational resources provided by Calcul Quebec and the Digital Research Alliance of Canada. Y.H. and L.P. acknowledge support from the Canada Research Chairs Program, the National Sciences and Engineering Council of Canada through grants RGPIN-2020-05073 and 05102, and the Fonds de recherche du Québec through grants 2022-NC-301305 and 300397. B.D.W. acknowledges support by the ANR BIG4 project, grant ANR-16-CE23-0002 of the French Agence Nationale de la Recherche; and the Labex ILP (reference ANR-10-LABX-63) part of the Idex SUPER, and received financial state aid managed by the Agence Nationale de la Recherche, as part of the programme Investissements d’avenir under the reference ANR-11-IDEX-0004-02. R.L. is supported by a generous donation by Eric and Wendy Schmidt with the recommendation of the Schmidt Futures Foundation.
R.L. thanks the Flatiron Institute for their hospitality and the Centre for Research in Astrophysics of Quebec for their support. P.L acknowledges support from the Simons Foundation. We thank Alexandre Adam for reading the manuscript and providing important feedback. 





\bibliographystyle{mnras}
\bibliography{bibliography} 




\appendix
\section{Additional Figures} \label{appendixA}

\begin{figure*}
 \begin{subfigure}{0.33\textwidth}
     \includegraphics[width=\textwidth]{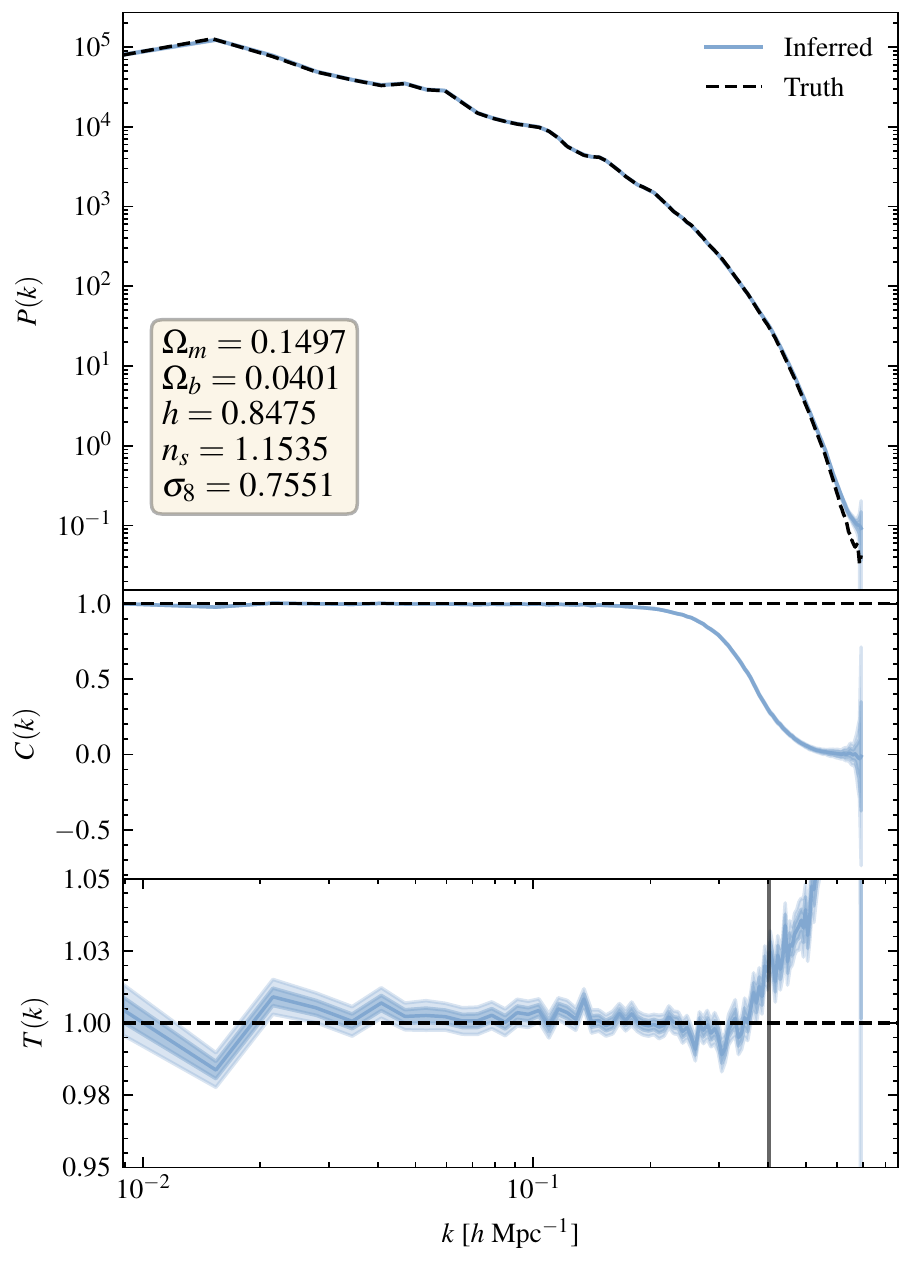}
     \label{fig:a}
 \end{subfigure}
 \hfill
 \begin{subfigure}{0.33\textwidth}
     \includegraphics[width=\textwidth]{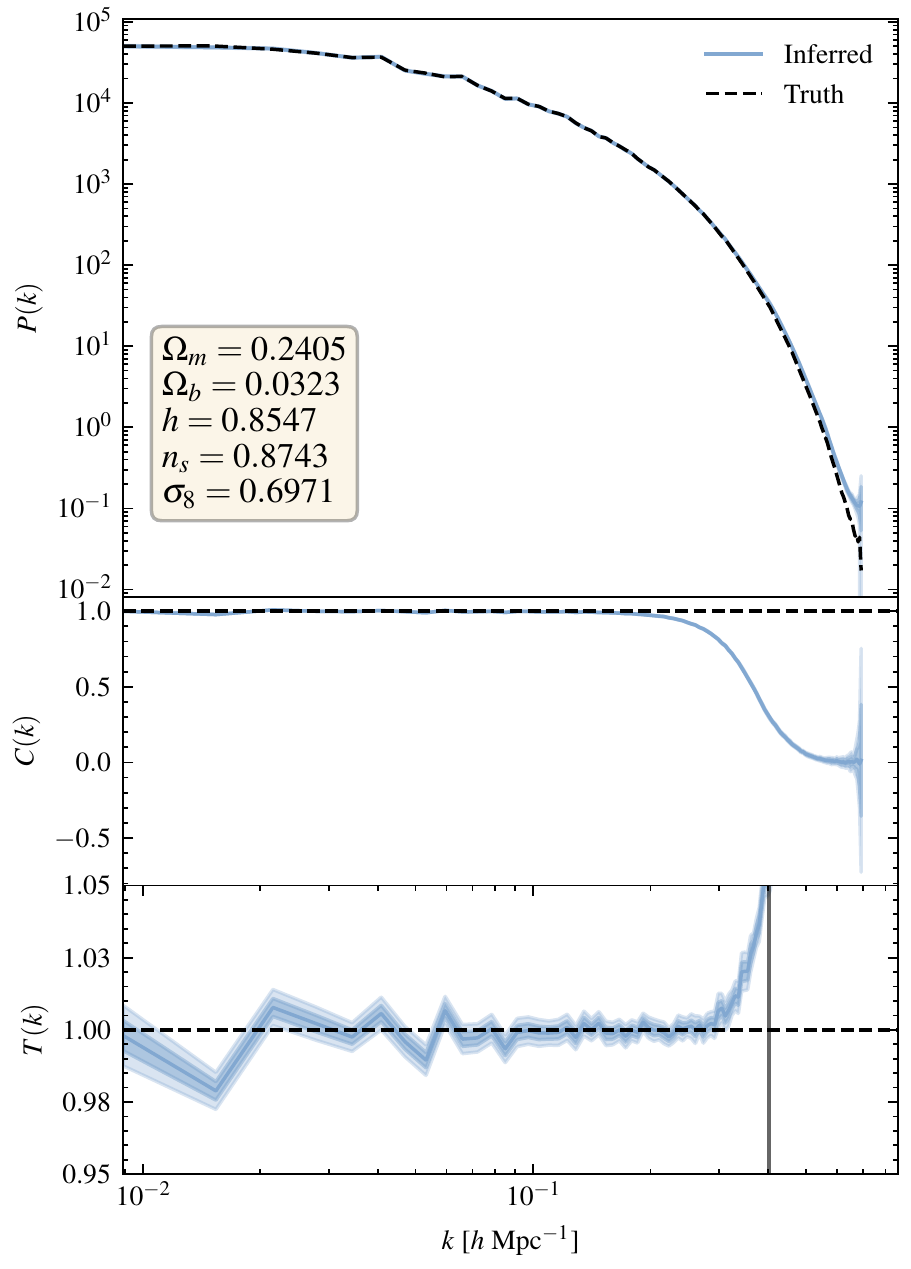}
     \label{fig:b}
 \end{subfigure}
 \hfill
 \begin{subfigure}{0.33\textwidth}
    \includegraphics[width=\textwidth]{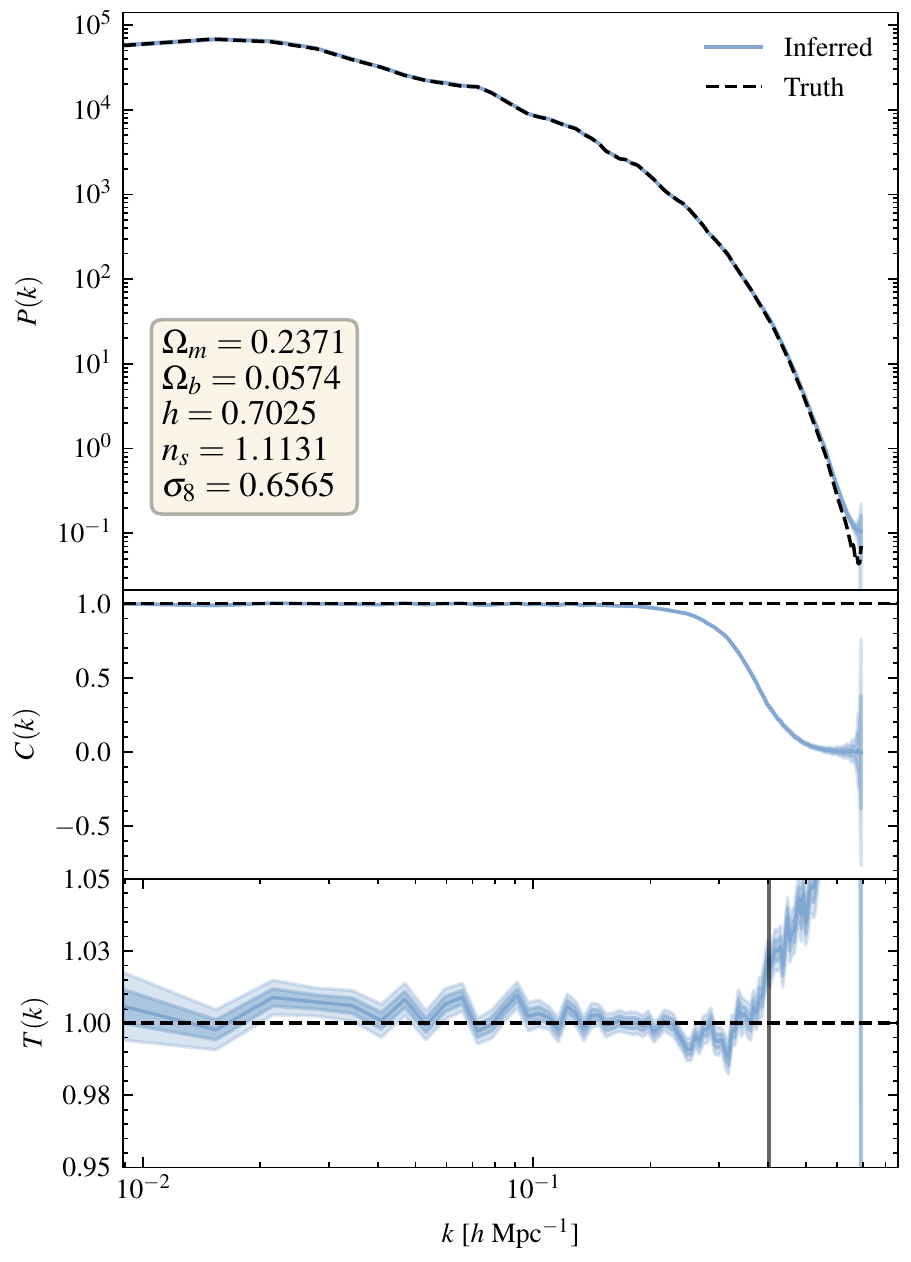}
     \label{fig:c}
 \end{subfigure}
 \medskip
 \begin{subfigure}{0.33\textwidth}
     \includegraphics[width=\textwidth]{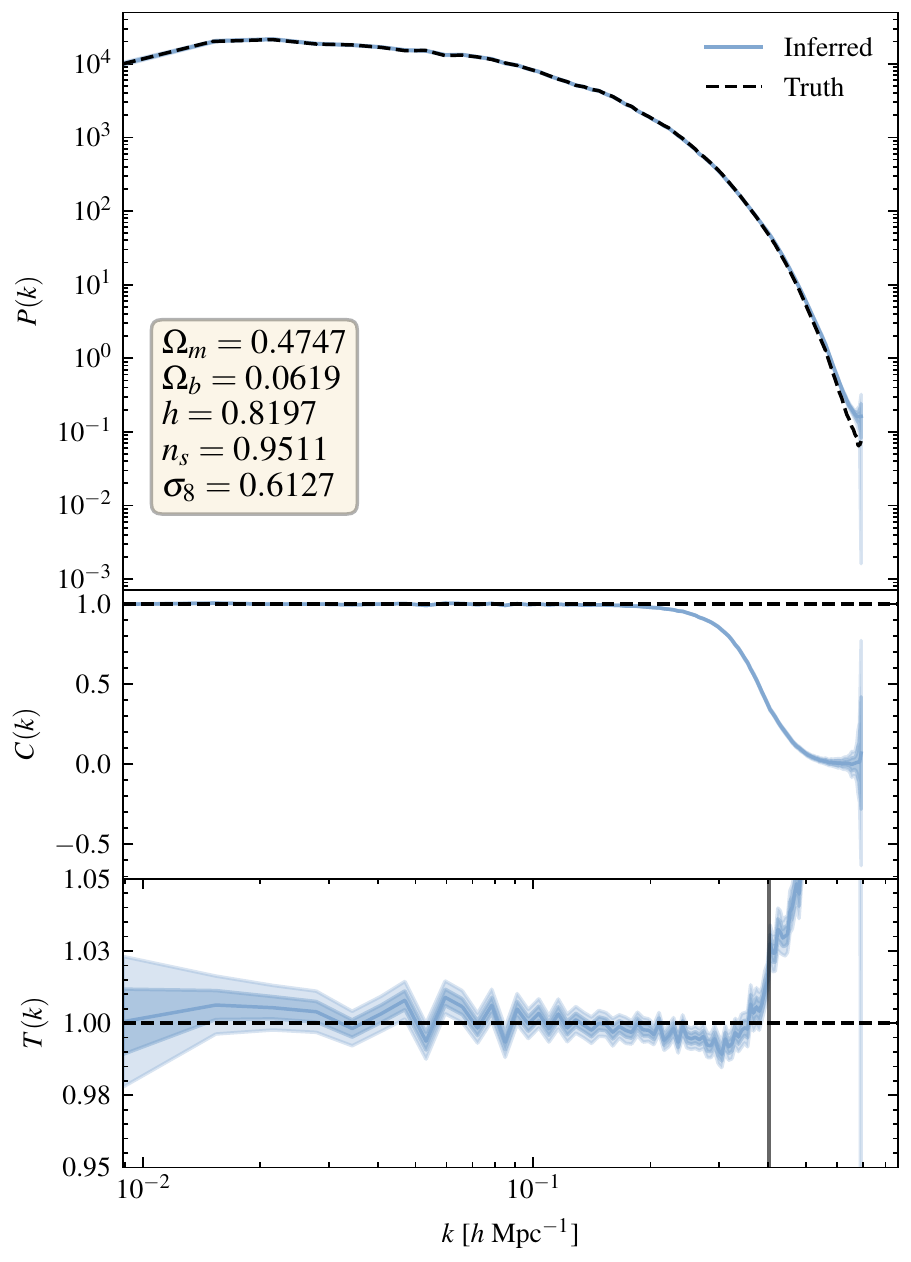}
     \label{fig:d}
 \end{subfigure}
  \hfill
 \begin{subfigure}{0.33\textwidth}
     \includegraphics[width=\textwidth]{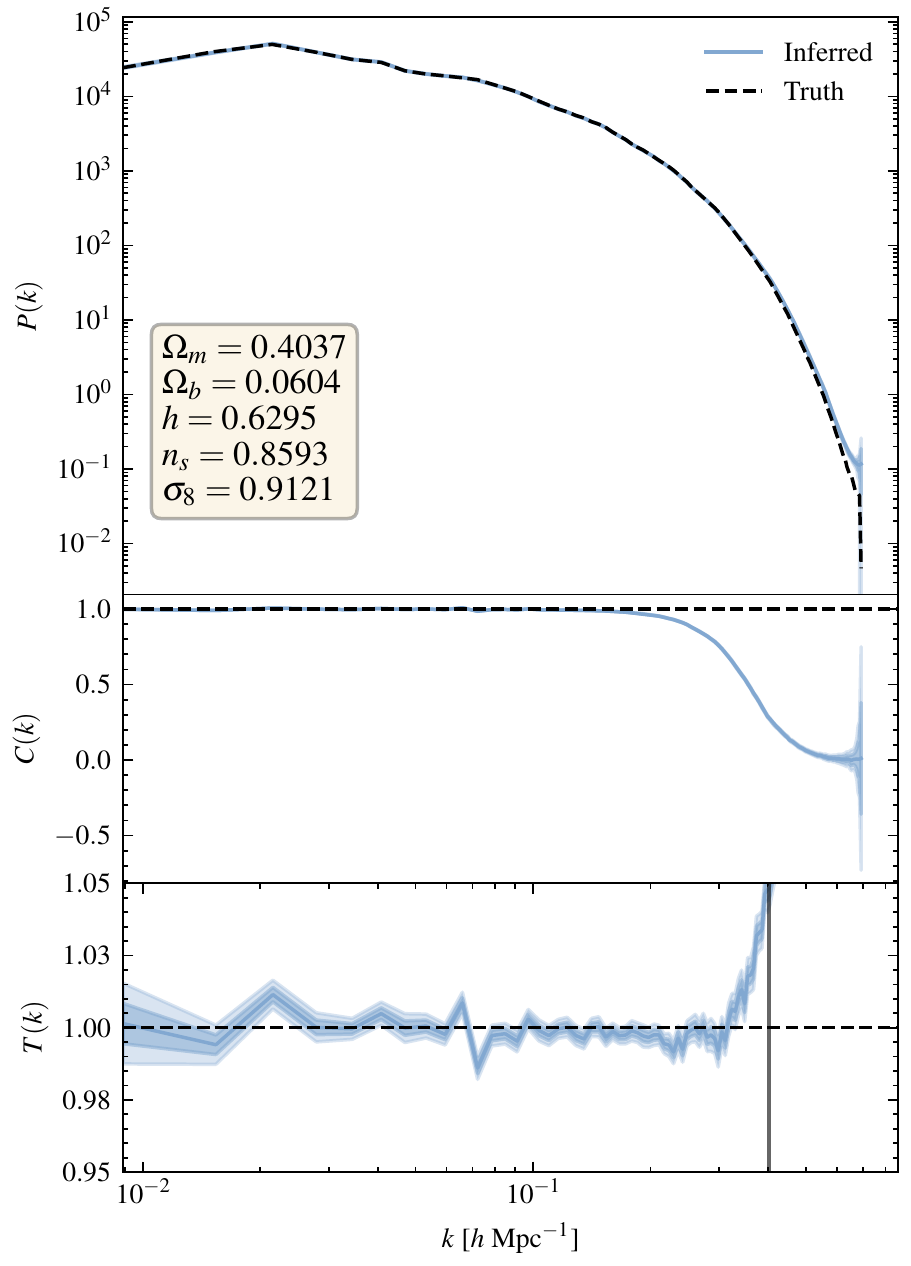}
     \label{fig:c}
 \end{subfigure}
 \hfill
 \begin{subfigure}{0.33\textwidth}
    \includegraphics[width=\textwidth]{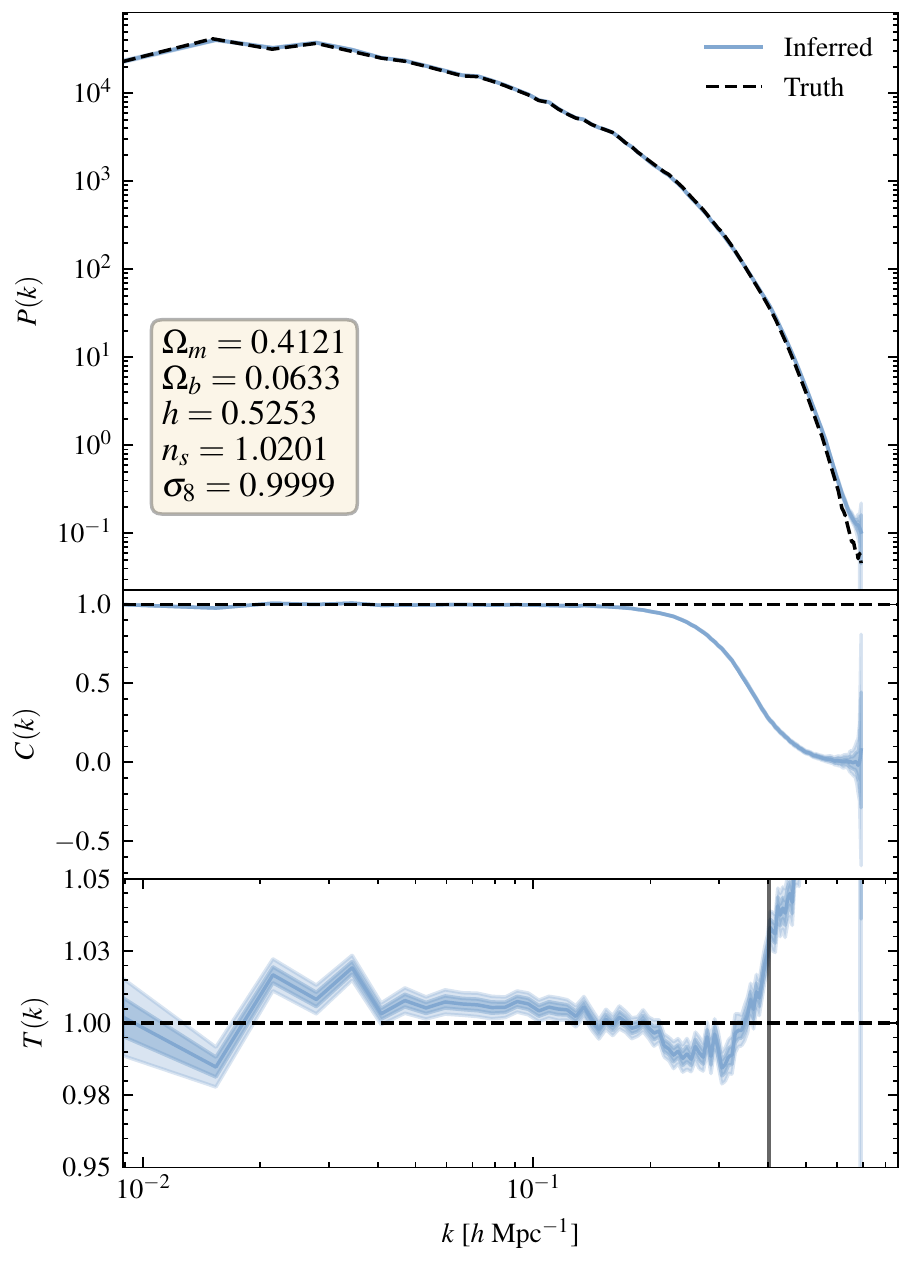}
     \label{fig:d}
 \end{subfigure}
  \caption{Summary statistics between generated posterior samples and ground truth initial conditions for N-body simulations runned with different values for the cosmological parameters. Note that the range of values for the cosmological parameters is much larger than the constraints from current observations. The results demonstrate that the score network is robust to different cosmologies and provides accurate reconstructions across different simulations.}
 \label{Label}
\end{figure*}

\begin{figure*}
 \begin{subfigure}{0.33\textwidth}
     \includegraphics[width=\textwidth]{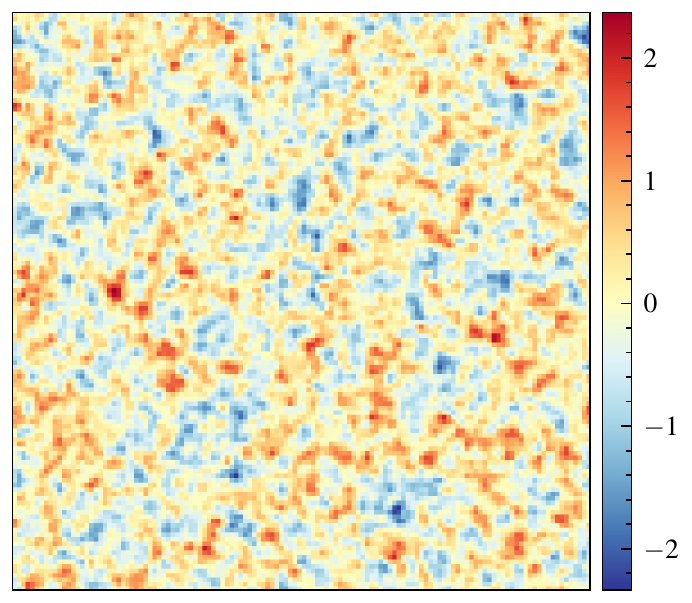}
     \label{fig:a}
 \end{subfigure}
 \hfill
 \begin{subfigure}{0.33\textwidth}
     \includegraphics[width=\textwidth]{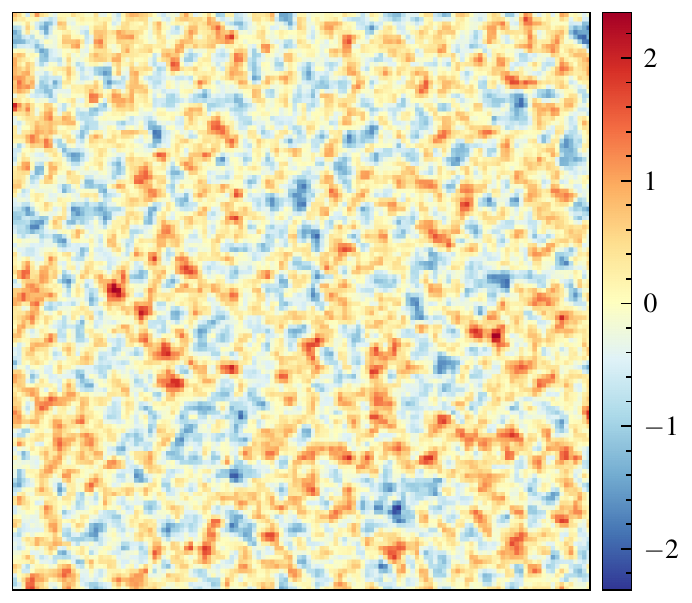}
     \label{fig:b}
 \end{subfigure}
 \hfill
 \begin{subfigure}{0.33\textwidth}
    \includegraphics[width=\textwidth]{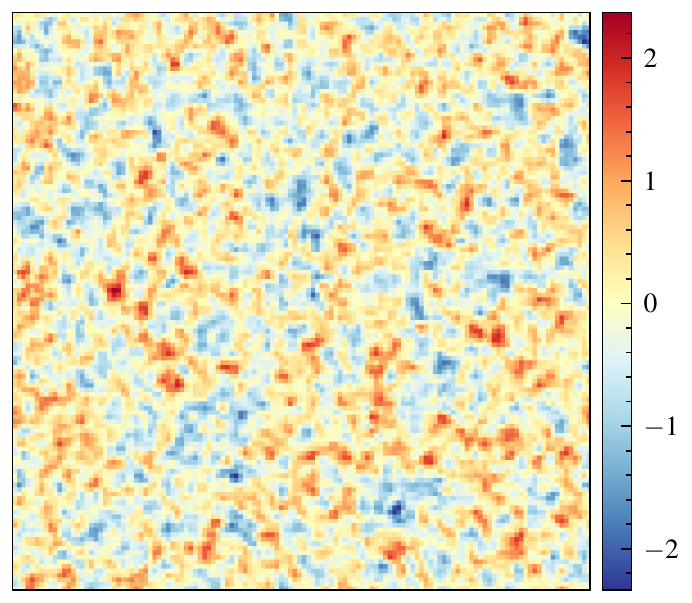}
     \label{fig:c}
 \end{subfigure}
 \medskip
 \begin{subfigure}{0.33\textwidth}
     \includegraphics[width=\textwidth]{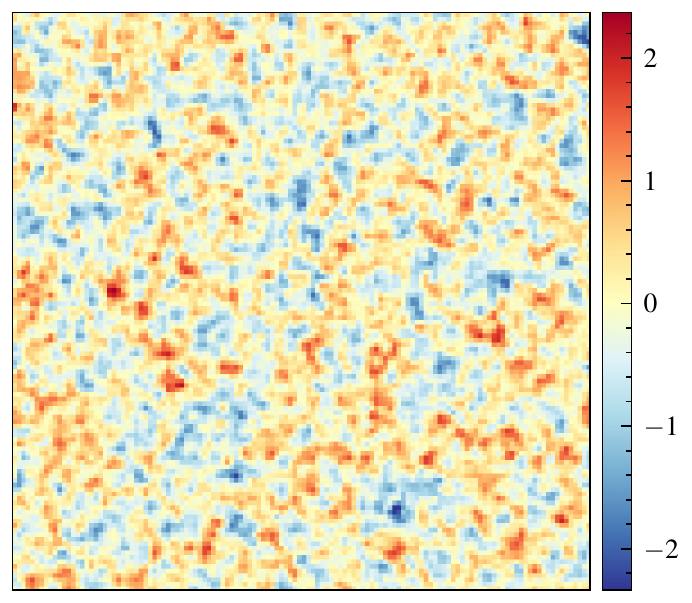}
     \label{fig:d}
 \end{subfigure}
  \hfill
 \begin{subfigure}{0.33\textwidth}
     \includegraphics[width=\textwidth]{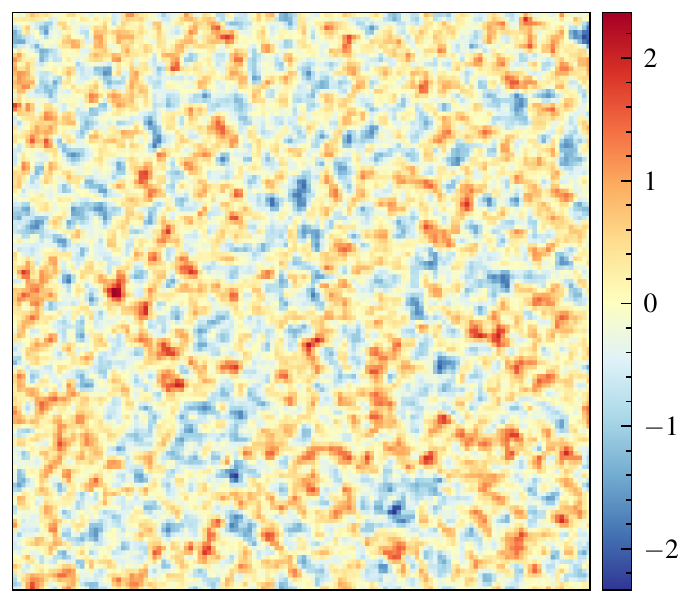}
     \label{fig:c}
 \end{subfigure}
 \hfill
 \begin{subfigure}{0.33\textwidth}
    \includegraphics[width=\textwidth]{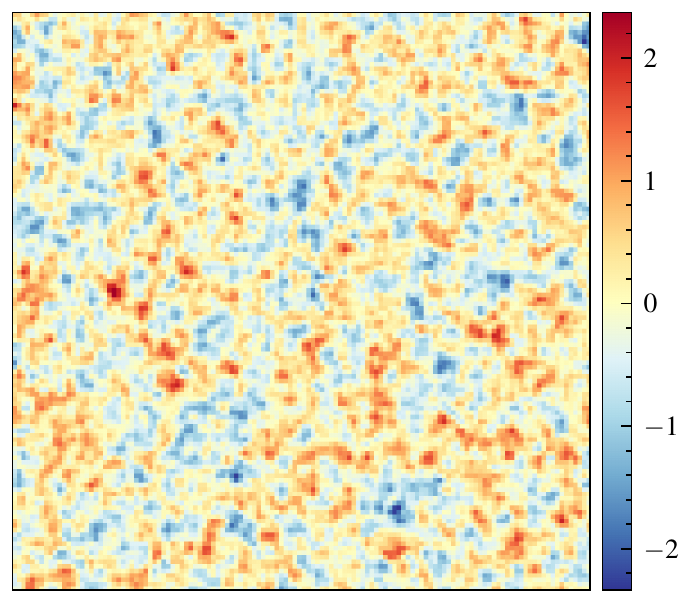}
     \label{fig:d}
 \end{subfigure}
  \caption{Independent samples of initial conditions sampled from the posterior $p(\bm{x}|\bm{y})$ based on the present-day density field $\bm{y}$ with fiducial cosmology from Figure \ref{obs_sample_true}. Each sample shown here accurately represents the true initial conditions, albeit with differences on the level of the posterior variance shown in Figure \ref{std_fid}.}
 \label{samples}
\end{figure*}


\bsp	
\label{lastpage}
\end{document}